\documentclass[twocolumn,english,aps,pra,showpacs,superscriptaddress]{revtex4}
\usepackage[T1]{fontenc}
\usepackage[latin9]{inputenc}
\usepackage{amsmath}
\usepackage{graphicx}
\makeatletter

\providecommand{\tabularnewline}{\\}

\@ifundefined{textcolor}{}
{%
 \definecolor{BLACK}{gray}{0}
 \definecolor{WHITE}{gray}{1}
 \definecolor{RED}{rgb}{1,0,0}
 \definecolor{GREEN}{rgb}{0,1,0}
 \definecolor{BLUE}{rgb}{0,0,1}
 \definecolor{CYAN}{cmyk}{1,0,0,0}
 \definecolor{MAGENTA}{cmyk}{0,1,0,0}
 \definecolor{YELLOW}{cmyk}{0,0,1,0}
 }


\newcommand{\ketbra}[2]{| #1 \rangle \langle #2 |}

\newcommand{\1}{{\rm 1\hspace{-0.9mm}l}}

\newcommand{\proof}{\noindent {\it Proof.\ }}

\newtheorem{fact}{Fact}

\newtheorem{theorem}{Theorem}
\newtheorem{lemma}{Lemma}
\newtheorem{definition}{Definition}

\newcommand{\stackidx}[4]{
\substack{
#1 #2 \\
#3 #4}
}

\makeatother

\usepackage{babel}
\begin{document}

\title{Collectibility for Mixed Quantum States }

\author{\L{}ukasz Rudnicki}

\email{rudnicki@cft.edu.pl}

\affiliation{Center for Theoretical Physics, Polish Academy of Sciences, Aleja
Lotnik{\'o}w 32/46, PL-02-668 Warsaw, Poland}

\author{Zbigniew Pucha\l{}a}

\affiliation{Institute of Theoretical and Applied Informatics, Polish Academy
of Sciences, Ba\l{}tycka 5, PL-44-100 Gliwice, Poland}

\affiliation{Smoluchowski Institute of Physics, Jagiellonian University, ul. Reymonta
4, PL-30-059 Krak{\'o}w, Poland}

\author{Pawe\l{} Horodecki}

\affiliation{Faculty of Applied Physics and Mathematics, Technical University
of Gda\'{n}sk, PL-80-952 Gda\'{n}sk, Poland}

\affiliation{National Quantum Information Centre of Gda\'{n}sk, PL-81-824 Sopot,
Poland}

\author{Karol \.{Z}yczkowski}

\affiliation{Smoluchowski Institute of Physics, Jagiellonian University, ul. Reymonta
4, PL-30-059 Krak{\'o}w, Poland}

\affiliation{Center for Theoretical Physics, Polish Academy of Sciences, Aleja
Lotnik{\'o}w 32/46, PL-02-668 Warsaw, Poland}

\date{October 25, 2012}

\begin{abstract}
Bounds analogous to entropic uncertainty relations
allow one to design practical tests to detect
quantum entanglement by a collective measurement 
performed on several copies of  the state analyzed.
This approach, initially worked out for pure states only [Phys. Rev. Lett. {\bf 107}, 150502 (2011)],
is extended here for mixed quantum states.
We define collectibility for any mixed states of a multipartite system. Deriving bounds for collectibility for positive partially transposed states 
of given purity provides a new insight into the structure of entangled quantum states.
In case of two qubits the application of complementary measurements 
and coincidence based detections leads to a new test of entanglement 
of pseudopure states.
\end{abstract}
\maketitle

\section{Introduction}

More than two decades ago the notion of the entanglement for mixed 
quantum states of a composite system
was worked out by Werner \cite{werner}.
Since then a lot of work has been done to 
develop efficient separability criteria 
and to design useful measures of quantum entanglement.
Although several possible quantities have been proposed and 
analyzed \cite{MCKB05,BZ06,PV07,HHHH09}
there is still a need for an efficient measure of quantum entanglement
which could be accessible in a 
real--life experiment \cite{WSDMB06,AL09,PR_Guhne}.

To get a full information about the analyzed quantum state
one can perform the scheme of quantum tomography, which allows one
to determine the degree of quantum entanglement \cite{SM+08}.
However,  the full scheme of quantum tomography requires
a large number of measurements, thus, it becomes not practical
for a higher dimensional systems. Therefore one can pose a question
how to get the maximal information about the degree of entanglement
of a given state performing relatively few measurements.

Some progress in this direction was achieved in \cite{Collectibility_RHZ},
in which a quantity based on collective measurements performed
on several copies of the system investigated was proposed. 
This quantity, called {\sl collectibility}, was defined for 
any pure state of a general composite quantum system, containing
an arbitrary number of $K$ subsystems, each describing $N$--level system.
Deriving inequalities analogous to entropic uncertainty relations
\cite{Deutsch,MU88} we established separability criteria based on collectibility.
To detect entanglement in the simplest two--qubit system 
we  proposed a four--photon experiment \cite{Collectibility_RHZ} based on 
Hong--Ou--Mandel interference \cite{HOM}.

%\medskip
%%%%%%%%%%%%%

Entanglement criteria based on pure--state collectibility are
reviewed in Section \ref{sec:pure}. The main goal of the present paper is to generalize the notion of collectibility 
and the related separability criteria to the general case of mixed
quantum  states. To this end we shall propose two strategies.

Firstly, in order to generalize the experimental part we shall modify
the pure--states separability criteria to take into account
contributions related to impurities of both copies of the investigated state.
We shall thus propose an entanglement test for ,,pseudopure'' states 
%can be considered as a one compromising between
%the number of observables measured
%and the number of interferences observed.
%In particular, it 
which employs the minimal number of
observables required for the scheme based on a single
Hong--Ou--Mandel interferometer. In general, this test seems
to be useful for quick but demanding tests of high quality sources. In addition it reports possible asymmetry in the character of the noise.
 In Section \ref{sec:experiment1} these modified
 entanglement criteria are presented
and the modified experimental setup is described. An extended analysis of the above results 
%entanglement criteria based on mixed--state collectibility 
is presented in Section \ref{sec:experiment2}.

In the second strategy, we shall generalize the definition of pure--state collectibility
and the corresponding  entanglement criteria to the case of an arbitrary mixed state
of a system consisting of $K$ subsystems, each supported on $N$ levels.
This is done in Section \ref{sec:mixed}. In Section \ref{sec:bipartiteWerner}
we investigate in more detail the bi--partite case.
%and propose an experimentally accessible uncertainty relation
%valid for mixed states of a two--qubit system.
Derivations of some formulae and proofs of certain lemmas 
are relegated to the Appendix.

\section{Pure--state collectibility}
\label{sec:pure}

\subsection{Maximal collectibility for multipartite pure states}

An entanglement test for pure states of $K$--quNit systems based
on uncertainty relations was proposed in \cite{Collectibility_RHZ}.
Consider a general case of a \textit{K--partite Hilbert space}
$\mathcal{H}=\mathcal{H}^{A}\otimes\mathcal{H}^{B}\ldots\otimes\mathcal{H}^{K}$
and for simplicity assume that $\dim\left(\mathcal{H}^{A}\right)=\ldots=\dim\left(\mathcal{H}^{K}\right)=N$.
In a first step we chose a set of $N$ separable pure states, $|\chi_{j}^{sep}\rangle=|a_{j}^{A}\rangle\otimes\ldots\otimes\left|a_{j}^{K}\right\rangle $,
where $|a_{j}^{I}\rangle\in\mathcal{H}^{I}$ and both indices run
$j=1,\dots,N$ and $I=A,\dots,K$. A crucial property of the states $|a_{j}^{I}\rangle$
 is that they are mutually orthogonal in each subspace, so that 
\begin{equation}
\left|a_{1}^{I}\right\rangle ,\ldots,\left|a_{N}^{I}\right\rangle \in\mathcal{H}^{I},\qquad\left.\left\langle a_{j}^{I}\right|a_{k}^{I}\right\rangle =\delta_{jk}.
\label{ortho}
\end{equation}

For a pure state of a composite system, $\left|\Psi\right\rangle \in\mathcal{H}$,
obeying the normalization condition $\langle\Psi|\Psi\rangle=1$,
we introduce: \begin{definition}\label{def1} Maximal collectibility
\cite{Collectibility_RHZ} of a pure state $\left|\Psi\right\rangle $
is 
\begin{equation}
Y^{{\rm max}}\left[\left|\Psi\right\rangle \right]=\max_{\left|\chi^{sep}\right\rangle }\prod_{j=1}^{N}\left|\!\left.\left\langle \Psi\right|\chi_{j}^{sep}\right\rangle \right|^{2}.
\label{MCp}
\end{equation}
\end{definition}To emphasize the fact that the above definition is
valid for pure states only 
we shall also call $Y^{{\rm max}}\left[\left|\Psi\right\rangle \right]$
the \textit{pure--state collectibility}. The maximum inside the formula
(\ref{MCp}) is necessary to assure an invariance with respect to
local unitary operations and shall be taken over the set of $N$ locally
mutually orthogonal pure states $|\chi^{sep}\rangle=\{|\chi_{1}^{sep}\rangle,\ldots,|\chi_{N}^{sep}\rangle\}$.

\subsection{Entanglement criteria based on pure--state collectibility}

In \cite{Collectibility_RHZ} we have shown that the pure--state collectibility
can serve as a simple entanglement test. Particularly, we have proven
two upper bounds for $Y^{\textrm{max}}\left[\left|\Psi\right\rangle \right]$.
The first one 
\begin{equation}
Y^{{\rm max}}\left[\left|\Psi\right\rangle \right]\leq N^{-N},\label{nieoznaczonosc}
\end{equation}
is valid for all states $\left|\Psi\right\rangle \in\mathcal{H}$,
while the second one 
\begin{equation}
Y^{{\rm max}}\left[\left|\Psi_{{\rm sep}}\right\rangle \right]\leq N^{-N\cdot K},\label{nieoznsep}
\end{equation}
must be satisfied if the state $\left|\Psi\right\rangle $ is a separable
state $\left|\Psi_{{\rm sep}}\right\rangle =\left|\Psi_{A}\right\rangle \otimes\ldots\otimes\left|\Psi_{K}\right\rangle $.
Since the second bound is much sharper than (\ref{nieoznaczonosc})
we obtain the following separability criteria
based on the pure--state collectibility  \cite{Collectibility_RHZ} 
\begin{equation}
\left(Y^{{\rm max}}\left[\left|\Psi\right\rangle \right]>\alpha_{K,N}\right)\Rightarrow\left(\left|\Psi\right\rangle \textrm{--- entangled}\right),\label{criteria}
\end{equation}
 where $\alpha_{K,N}=N^{-N\cdot K}$ plays the role of a discrimination
parameter.

\subsection{Experimentally accessible criteria for a pure state of many qubits}

In the case of $K$--qubits we are able to perform analytically a
first step of the maximization procedure, i.e. to maximize over a
pair of two vectors $\left|a_{1}^{A}\right\rangle ,\left|a_{2}^{A}\right\rangle \in\mathcal{H}^{A}$
belonging to the first Hilbert subspace. We obtain an expression
for the \textit{collectibility} of a pure state \cite{Collectibility_RHZ}:
\begin{eqnarray}
Y_{a}\left[\left|\Psi\right\rangle \right] & = & \max_{\left|a_{1}^{A}\right\rangle ,\left|a_{2}^{A}\right\rangle }\prod_{j=1}^{N}\left|\!\left.\left\langle \Psi\right|\chi_{j}^{sep}\right\rangle \right|^{2}\label{Collectibility}\\
 & = & \frac{1}{4}\left(\sqrt{G_{11}G_{22}}+\sqrt{G_{11}G_{22}-\left|G_{12}\right|^{2}}\right)^{2}\!\!\!,\nonumber 
\end{eqnarray}
where the coefficients $G_{jk}=\langle\varphi_{j}|\varphi_{k}\rangle$
are elements of the\textbf{ }Gram matrix for two ($j=1,2$) vectors
$|\varphi_{j}\rangle=\left(\left\langle a_{j}^{B}\right||\otimes\ldots\otimes\left\langle a_{j}^{K}\right|\right)|\Psi\rangle\in{\cal H}^{A}$.
The vectors $|\varphi_{j}\rangle$ represent the state $\left|\Psi\right\rangle $
projected onto two orthogonal, separable states $\left|a_{j}^{B}\right\rangle \otimes\ldots\otimes\left|a_{j}^{K}\right\rangle $
of $K-1$ qubits.

A particular case of the separability criterion (\ref{criteria}) 
for the $K$--qubit system reads 
\begin{equation}
\left(Y_{a}\left[\left|\Psi\right\rangle \right]>\alpha_{K,2}\right)\Rightarrow\left(\left|\Psi\right\rangle \textrm{--- entangled}\right).\label{criteria_Col}
\end{equation}
An important advantage of the above separability criterion based on 
collectibility is its possible experimental implementation.
 In \cite{Collectibility_RHZ} we proposed an experiment
based on Hong--Ou--Mandel (H--O--M) \cite{HOM} interferometry, in
which all coefficients $G_{jk}$ can be measured if two copies of
a two--qubit pure state are given. If $Y_{a}\left[\left|\Psi\right\rangle _{AB}\right]>1/16$
then the two--qubit state $\left|\Psi\right\rangle _{AB}$ is entangled.
In the following sections of this work
we will generalize this approach for the mixed states of 
a multipartite system.

\section{Two qubits in a mixed state --- pseudopure entanglement test by controlling remote purity}\label{sec:experiment1}

\begin{figure}
\centering\includegraphics[scale=0.35]{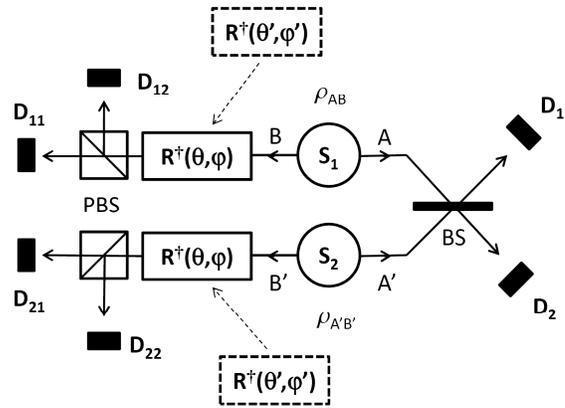} 
\caption{
Determination of the Gram matrix via conditional
overlapping in the case of two polarization--entangled photon
pairs. Each source produces a pair of photons in a 
general mixed--polarization state  $\rho_{AB}$.
The statistics of pairs of clicks is collected on the left hand side
after two elements are measured.
On the right side the H--O--M interference is performed.
The square of the Gram matrix element equals to
the probability of the pair of clicks multiplied by that of
double click on the right. The extra boxes with the primed arguments of the rotations $R(\theta',\phi')$ 
represent the measurements in a basis complementary to the 
original one associated with the rotations $R(\theta,\phi)$
(see the main text). The hermitian conjugate at the rotations
correspond to the dualism of Schr{\"o}dinger and Heisenberg pictures.
}
%%%%%%%%%%%
%Determination of the Gram matrix via conditional overlapping in the
%case of two polarization--entangled photon pairs. Each source produces
%a pair of photons in a --- in general mixed --- polarization state
%$\rho_{AB}$. On the left side the statistics of pairs of clicks after
%two PBS--elements are measured. On the right side the H--O--M interference
%is performed. The square of the Gram matrix element equals to the
%probability of the pair of the clicks multiplied by that of double
%click on the right. 
\label{Fig.2} 
\end{figure}

The starting point of our considerations shall be the original experimental
setup \cite{Collectibility_RHZ} with the pure state $\left|\Psi\right\rangle _{AB}$
substituted by the mixed state $\rho_{AB}$ shared by Alice and Bob.
Apart from two sources of the same copy of the state it involves 
 the 50:50 beam splitter (BS), two polarization
rotators $R^{\dagger}(\theta,\phi)$ in the same setting and the polarized
beam splitters (PBS)  --- see Fig. \ref{Fig.2}. 
Let us define $p_{ij}(+,+)$ as the probability
of double click after the beam splitter. We also denote by $p_{1i}\equiv p_{1}\left((-1)^{i+1}\right)$
($p_{2i}\equiv p_{2}\left((-1)^{i+1}\right)$) the probability of
click in the $D_{1,i}$-th detector ($D_{2,i}$-th detector), i.e.
one of the detectors located after upper PBS (lower PBS). The Gram
matrix elements result in: 
\begin{equation}
\left|G_{ij}\right|^{2}=p_{1i}p_{2j}\left(1-2p_{ij}(+,+)\right).
\end{equation}

Note now that whenever Alice gets the two results in the same index
(which corresponds to the probabilities $p_{1i}$, $p_{2i}$ for $i=1,2$)
then she remotely produces at Bob site a pair of the same state, say
$\sigma_{+}$ --- with probability $p_{11}=p_{21}\equiv p_{+}$ ---
or $\sigma_{-}$ with probability $p_{12}=p_{22}\equiv p_{-}$ respectively,
which further subjects to H--O--M interference. When however she gets
the results with different second index, namely either $p_{11}\equiv p_{+}$,
$p_{22}\equiv p_{-}$ or $p_{12}\equiv p_{-}$, $p_{21}\equiv p_{+}$
then she produces a pair of different states, i.e. $\sigma_{+}$,
$\sigma_{-}$ or $\sigma_{-}$, $\sigma_{+}$ which will come into
the H--O--M interferometer on the right hand side. This means that
finally we have the following relation between the measurable quantities
($G_{\pm\pm}$ and $G_{\pm\mp}$) and the mathematical ones ($G_{ij}$,
$i,j=1,2$): $G_{11}\equiv G_{++}$, $\left|G_{12}\right|\equiv\left|G_{+-}\right|$,
$\left|G_{21}\right|\equiv\left|G_{-+}\right|$ and $G_{22}\equiv G_{--}$,
where: \begin{subequations}
\begin{equation}
G_{++}=p_{+}\sqrt{\textrm{Tr}\left(\sigma_{+}^{2}\right)},\quad G_{--}=p_{-}\sqrt{\textrm{Tr}\left(\sigma_{-}^{2}\right)},\label{DefG1}
\end{equation}
and  
\begin{equation}
\left|G_{+-}\right|^{2}=p_{+}p_{-}\textrm{Tr}\left(\sigma_{+}\sigma_{-}\right)=\left|G_{-+}\right|^{2}.\label{DefG2}
\end{equation}
Some other \end{subequations} quantities will also be used 
in our formulas: 
\begin{eqnarray}
\textrm{Tr}\left(\sigma_{+}\sigma_{-}\right) & = & 1-2p_{12}(+,+),\nonumber \\
\textrm{Tr}\left(\sigma_{+}^{2}\right) & = & 1-2p_{11}(+,+),\nonumber \\
\textrm{Tr}\left(\sigma_{-}^{2}\right) & = & 1-2p_{22}(+,+).\label{overloaps}
\end{eqnarray}
We may expect that one has $p_{12}(+,+)=p_{21}(+,+)$
up to measurement accuracy since the two copies of the 
state $\rho_{AB}$ are assumed to be the same.

\subsection{Modified version of the experiment involving two complementary observables}
Consider the case in which Alice and Bob share a two--qubit state
$\rho_{AB}$ and Alice performs measurements of \textit{two} complementary
binary observables --- $\hat{n}\vec{\sigma}$ and $\hat{n}'\vec{\sigma}'$.
Below we shall refer to the axes $\hat{x}$ and $\hat{z}$ but it
is only due to simplicity of the derivation, which is in fact invariant
under the choice of the orthogonal pair $\hat{n}$ and $\hat{n}'$. 
Under each of the four results Alice produces remotely one of the four states
on Bob side, $\sigma_{\pm}$ (when she gets the result $\pm$ while
measuring $\hat{n}$) or $\sigma_{\pm}'$ (when she gets the result
$\pm$ while measuring $\hat{n}'$).

The experimental setup enclosed on Fig. \ref{Fig.2} is almost
the same as the one designed for 
the original inequality $Y_{a}\left[\left|\Psi\right\rangle _{AB}\right]>1/16$.
The only difference is that while in the original scheme one performs 
in both arms a single measurement (determined by some specific choice of the
unitary rotation $R(\theta,\phi)$), here we perform the same
experiment for two selected rotations $R(\theta,\phi)$ and $R(\theta',\phi')$
labeled by a pair of angles defining two orthonormal versors $\hat{n}$
and $\hat{n}'$.

We shall measure the degree of purity  of $\sigma_{\pm}$ and $\sigma_{\pm}'$
(which is directly measurable according to (\ref{overloaps})) by
four parameters $\epsilon_{\pm}$, $\epsilon_{\pm}'$ as follows:\begin{subequations}
\begin{equation}
\textrm{Tr}\left(\sigma_{\pm}^{2}\right)\geq1-\epsilon_{\pm},\label{remote-purity}
\end{equation}
\begin{equation}
\textrm{Tr}\left(\left(\sigma_{\pm}'\right)^{2}\right)\geq1-\epsilon_{\pm}'.\label{remote-purity1}
\end{equation}
\end{subequations} Moreover, we have other parameters which are the
probabilities that Alice produces remotely the states $\sigma_{\pm}$
($\sigma_{\pm}'$) which we have denoted by $p_{\pm}$ ($p_{\pm}'$).
The final parameters we need are the overlaps 
%$\textrm{Tr}\left(\sigma_{+}\sigma_{-}\right)$
%and $\textrm{Tr}\left(\sigma_{+}'\sigma_{-}'\right)$ 
between the "$\sigma$" states, which are easily
measurable as one can see from (\ref{overloaps}). The probabilities,
purity parameters and the overlaps are the only data from the experiment.
While finally in practice we shall put equalities in the formulas
(\ref{remote-purity}, \ref{remote-purity1}) we keep inequalities
to stress that this may take into account all experimental statistical
errors resulting from data analysis. The test works well provided 
all these four states produced remotely at the Bob site are characterized by a
high purity, which resembles the ideal situation. 
If the initial state $\rho_{AB}$
was a pure state, then all four states $\sigma_{\pm}$, $\sigma_{\pm}'$
would be completely pure. That is why we refer to the test as \textit{pseudopure}
but it is fully general, i.e. we do not need to have any additional
assumption on the structure of the state. In order to generalize the
entanglement criteria $Y_{a}\left[\left|\Psi\right\rangle _{AB}\right]>1/16$
to the case of the two--qubit mixed state $\rho_{AB}$ we shall establish
the following theorem:
\begin{theorem}
\label{th:bound1} Any separable
two--qubit state $\rho_{AB}$ satisfies the following inequality:
\begin{equation}
G_{++}G_{--}-\left|G_{+-}\right|^{2}\leq
\frac{\eta+\left(G_{++}+G_{--}\right)^{2}-1}{2},
\label{corrected1}
\end{equation}
where 
\begin{equation}
\label{etat}
\eta\equiv\eta\left(\epsilon_{+}\epsilon_{-};p_{+}p_{-};\epsilon'\right)
=8p_{+}p_{-}\sqrt{\epsilon_{+}\epsilon_{-}}+2\epsilon'\leq2,
\end{equation}
and \textit{$\epsilon'=\max\left[\epsilon_{+}',\epsilon_{-}'\right]$.
}\end{theorem} The proof of the above theorem together with an extended
discussion of the inequality (\ref{corrected1}) can be found in Section
\ref{sec:experiment2}. As a weaker consequence of Theorem \ref{th:bound1}
we obtain the separability criteria in terms of the collectibility
$Y_{a}$ and the parameter  $\eta$  given in (\ref{etat}):
\begin{equation}
\left(\! Y_{a}\left[\rho_{AB}\right]>\frac{1}{16}+\frac{1}{4}\left(\frac{\eta}{2}+\!\sqrt{\frac{\eta}{2}}\right)\!\!\right)\Rightarrow\left(\rho_{AB}\;\textrm{--- entangled}\right).\label{criteria_Col2}
\end{equation}

%%%%%%%%%%%%%%%%%%%%%%%

%Let us now  consider the special case 
%in which the source produces partially depolarized singlet states
%called a two-qubit Werner state:
%\begin{equation}
%\rho_{AB}(p)=(1-p) |\Psi_+\rangle \langle \Psi_+| + p \frac{I \otimes I}{4}
%\label{dep-singlet}
%\end{equation} 
%The assumption about uniform noise is a very natural practice in context of PDC 
%experiments with entanglement. Furthermore, the general 
%form of the state (\ref{dep-singlet}) can be enforced by local random unitaries of the form 
%$U \otimes U$ which can be applied via manipulation with half-wave plates.
Note that we may rewrite the formula (\ref{corrected1}) as a nonlinear entanglement witness value
\begin{equation}
{\cal W} (\rho)=\frac{1}{2}(\eta + G_{++}^2 + G_{--}^{2} +2 \left|G_{+-}\right|^{2} - 1) \geq 0. 
\label{corrected2}
\end{equation}
We have checked the above test for the Werner 
states with admixture of uniform noise with probability $p$:
 \begin{equation}
\rho_{AB}(p)=(1-p) |\Psi_+\rangle \langle \Psi_+| + p \frac{I \otimes I}{4}.
\label{dep-singlet}
\end{equation} 
It is known that the threshold for separability
is $p=2/3=0.(6)$, while the scheme with two Hong--Ou--Mandel
interferometers \cite{BCEHAS05} 
(which realizes the purity/entropy separability test \cite{HHH96})
gives the threshold as $p=1-1/\sqrt{3}\approx 0.422$6. 
The pseudomixture scheme proposed here provides a smaller value $p=1-\sqrt{3}/2\approx 0.1180$.
This implies that the pseudomixed test is essentially 
dedicated to high quality sources,
i.e. small perturbations of purity. It allows
to test them faster than in  \cite{BCEHAS05} 
(provided that the source is good enough)
as it requires two settings on Alice side and the interference on Bob side
as opposed to usual tests where two Hong--Ou--Mandel interferences are used.
In a sense it offers a balanced compromise between the number 
of observables measured and the number of interferences performed.

\section{Mixed--state collectibility for multipartite systems}
 \label{sec:mixed}

   A first natural effort to generalize the definition of the pure--state
collectibility could be to rewrite the right hand side of the definition
(\ref{MCp}) in the following form
\begin{equation}
\max_{\left|\chi^{sep}\right\rangle }\prod_{j=1}^{N}
\langle \chi_{j}^{sep}\left|\Psi\right\rangle \langle \Psi\left|\chi_{j}^{sep}\right\rangle .
\label{right hand side}
\end{equation}
One could suspect, that a simple modification $\left|\Psi\right\rangle \left\langle \Psi\right|\mapsto\rho$
will make  the above formula suitable for an arbitrary mixed
state $\rho$. However, we require the \textit{mixed--state collectibility}
to be capable to identify entanglement. On the other hand the 
separable maximally mixed state of rank $N$
can be represented as a sum of $N$ 
mutually orthogonal separable states $|\chi^{sep}\rangle$,
\begin{equation}
\rho_{\chi}=\frac{1}{N}\sum_{k=1}^{N}\left|\chi_{k}^{sep}\right\rangle \left\langle \chi_{k}^{sep}\right|,\label{maxMixedRankN}
\end{equation}
%related to the $N$--dimensional subspace of $\mathcal{H}$ spanned
%by the set $|\chi^{sep}\rangle$ consisting of $N$ orthogonal vectors,
This implies that  formula (\ref{right hand side})
 with $\left|\Psi\right\rangle \left\langle \Psi\right|$ replaced by 
$\rho_{\chi}$ immediately gives the value $N^{-N}$, which is the upper bound (\ref{nieoznaczonosc}).
The same bound is attained by any pure, maximally entangled state. 
% But the maximally mixed state $\rho_{\chi}$ is separable. 
The above observation
shows that the quantity based on expression (\ref{right hand side})
does not allow us to distinguish between the maximally entangled pure state
and the maximally mixed separable state.

 In order to remove this ambiguity we introduce: 
\begin{definition} The mixed--state collectibility
of a mixed state $\rho$ is 
\begin{equation}
Y^{{\rm max}}\left[\rho\right]=\left(\max_{\left|\chi^{sep}\right\rangle }\prod_{j,k=1}^{N}\left\langle \chi_{j}^{sep}\right|\rho\left|\chi_{k}^{sep}\right\rangle \right)^{1/N}.
\label{MCm}
\end{equation}
\end{definition}Since we also take into account the \textit{off--diagonal}
terms of the density matrix, we necessarily obtain
for the maximally mixed state
 $Y^{{\rm max}}\left[\rho_{\chi}\right]=0$,
because the density matrix  $\rho_{\chi}$ is diagonal.

Note, that for the mixed--state collectibility we use
the same symbol $Y^{{\rm max}}\left[\cdot\right]$ 
as for the pure--state collectibility, 
thus only the argument ($\left|\Psi\right\rangle $
or $\rho$) allows one to distinguish the difference. In fact, we
can use the same symbol for both quantities, since the $N$--th root
in the definition (\ref{MCm}) assures that 
$Y^{{\rm max}}\left[\left|\Psi\right\rangle \left\langle \Psi\right|\right]=Y^{{\rm max}}\left[\left|\Psi\right\rangle \right]$.
The mixed--state collectibility (\ref{MCm}) calculated for a rank--one
density operator $\rho=\left|\Psi\right\rangle \left\langle \Psi\right|$
is equal to the earlier defined pure--states collectibility (\ref{MCp}),
what makes both definitions consistent.

\subsection{Criteria based on mixed--state collectibility}

After we have defined the mixed--state collectibility we would like
to generalize the upper bounds (\ref{nieoznaczonosc}) and (\ref{nieoznsep}).
To this end let us consider a $D\times D$ hermitian, positive semi--definite
matrix $\rho\geq0$, $\rho=\rho^{\dagger}$, with fixed trace $\textrm{Tr}\rho=\textrm{const}$
and purity $\textrm{Tr}\left(\rho^{2}\right)=\textrm{const}$. Let
us denote by $\rho_{ij}$, $i,j=1,\ldots,D$ the matrix elements of
$\rho$. We shall establish the following lemma\begin{lemma}\label{lemma1}For
every $N\leq D$ we have\begin{subequations}
\begin{equation}
\left(\prod_{i,j=1}^{N}\!\!\rho_{ij}\right)^{\!\!1/N}\!\!\!\!\!\!\!\leq r_{N}\left(D,\textrm{Tr}\rho,\textrm{Tr}\left(\rho^{2}\right)\right),\label{lem1Eq1}
\end{equation}
where:
\begin{equation}
r_{N}=\left(\frac{\textrm{Tr}\rho}{N}\right)^{N}\left(1-D\Phi\right)^{\frac{N-1}{2}}\left(1-\tilde{D}\Phi\right)^{\frac{N+1}{2}}\!\!,\label{lem1Eq2}
\end{equation}
\begin{equation}
\Phi=\frac{D+\tilde{D}-1-\sqrt{\left(D+\tilde{D}-1\right)^{2}+4D\tilde{D}\left(\xi-1\right)}}{2D\tilde{D}},
\end{equation}
\begin{equation}
\xi=\frac{\textrm{Tr}\left(\rho^{2}\right)}{\left(\textrm{Tr}\rho\right)^{2}}\in\left[D^{-1},1\right],\qquad \tilde{D}=D-N.\label{lem1Eq3}
\end{equation}
\end{subequations}
\end{lemma}
Proof of Lemma \ref{lemma1} can be
found in Appendix \ref{App:product}.

A mixed--states analogue of the general upper bound (\ref{nieoznaczonosc})
reads
\begin{equation}
Y^{{\rm max}}\left[\rho\right]\leq r_{N}\left(N^{K},1,\textrm{Tr}\left(\rho^{2}\right)\right),
\label{nieoznaczonosc_mixed}
\end{equation}
because in our particular case we have $D=N^{K}$ and $\textrm{Tr}\rho=1$.
One can check that for pure states we obtain $r_{N}\left(D,1,1\right)=N^{-N}$
independently of the dimension $D$ of a composite Hilbert space,
so that the bound (\ref{nieoznaczonosc}) is recovered. Moreover,
the bound (\ref{nieoznaczonosc_mixed}) is an increasing function
of purity $\textrm{Tr}\left(\rho^{2}\right)$ with the minimal value
equal to $0$ for the maximally mixed state of rank $N^{K}$ and the
maximal value reached for pure states.

The task to generalize the upper bound (\ref{nieoznsep}) and the
entanglement criteria (\ref{criteria}) is more difficult,
 since the conditions
for separability of mixed states are more complex, comparing with
the case of pure states. In order to deal with the bipartite case
we shall prove the following statement:\begin{theorem}\label{theorem 1}Assume
that $K=2$, and $\mathcal{H}_{2}=\mathcal{H}^{A}\otimes\mathcal{H}^{B}$.
If $\rho$ acting on $\mathcal{H}_{2}$ is PPT (has positive partial
transpose), so that $\rho^{T_{A}}\geq0$ where $T_{A}=T\otimes\1$,
then 
\begin{equation}
Y^{\max}\left[\rho\right]\leq N^{-2N}.\label{PPT inequality}
\end{equation}

\end{theorem}\proof Denote $\rho_{\stackidx{j}{j}{k}{k}}=\left\langle \chi_{j}^{sep}\right|\rho\left|\chi_{k}^{sep}\right\rangle $.
The partial transposition $(\cdot)^{T_{A}}$, transforms indices as,
\begin{equation}
(\cdot)^{T_{A}}:\stackidx{m}{\mu}{n}{\nu}\mapsto\stackidx{n}{\mu}{m}{\nu}.
\end{equation}
After the partial transposition the state is non negative, what implies
that \cite{matrix analysis} 
\begin{equation}
\left|\rho_{\stackidx{j}{j}{k}{k}}\right|^{2}\leq\rho_{\stackidx{j}{k}{j}{k}}\rho_{\stackidx{k}{j}{k}{j}}.\label{PPT-positivity}
\end{equation}
The above inequalities are strongly related with the separability
criteria derived in \cite{flo,kryteriadod}. Making use
 of the property (\ref{PPT-positivity}) we get 
\begin{equation}
Y^{\max}\left[\rho\right]\leq\left(\prod_{j,k=1}^{N}\rho_{\stackidx{j}{k}{j}{k}}\right)^{1/N}.\label{partial inequality}
\end{equation}
All factors appearing in the last product are diagonal elements of
$\rho$. Moreover, $\rho$ has exactly $N^{2}$ diagonal elements,
thus all of them occur in the product inside (\ref{partial inequality}).
As the last step we shall use the inequality between arithmetic and
geometric means and derive the result (\ref{PPT inequality}) 
\begin{equation}
Y^{\max}\left[\rho\right]\leq\left(\prod_{j,k=1}^{N}\rho_{\stackidx{j}{k}{j}{k}}\right)^{1/N}\!\!\!\!\!\leq\left(\frac{1}{N^{2}}\sum_{j,k=1}^{N}\rho_{\stackidx{j}{k}{j}{k}}\right)^{N}\!\!=N^{-2N},
\end{equation}
where in the last equality we used the property that $\textrm{Tr}\rho=1$. 

Consider now, the case of $K$--qubits, i.e. $N=2$. Using the same
method as before it can be proven that:\begin{fact}\label{fact1}
If $N=2$ and $\rho$ is PPT for partial transpositions with respect
to all possible bi--partite splittings of the $K$--partite system,
then 
\begin{equation}
Y^{\max}\left[\rho\right]\leq\frac{1}{16\left(2^{K-1}-1\right)}.\label{PPT inequality2}
\end{equation}

\end{fact}Inequality (\ref{PPT inequality2}) is sharp (can be saturated)
for all values of $K$, and in the case of two qubits coincides with
(\ref{PPT inequality}) for $N=2$. 

Using the inequalities (\ref{PPT inequality}) and (\ref{PPT inequality2})
we are able to derive counterparts of the criteria (\ref{criteria})
for two cases of mixed quantum states: a) $K=2$ and arbitrary $N$
and, b) $N=2$ and arbitrary $K$, with a restriction, that we detect
whether the state is NPPT (not PPT) instead of being not separable.

\subsection{An efficiency of NPPT states detection}

An answer to the question concerning the efficiency of the
entanglement criteria considered 
 depends strongly on the investigated state.
In particular, there are states which are too close to the set of
separable (or PPT) states, to be detected by given criteria. In this
paragraph, we shall characterize the set of NPPT states which is covered
by the entanglement test based on mixed--state collectibility.

At first we note that
 the purity $\textrm{Tr}\left(\rho^{2}\right)$ is bounded 
from below by the value $\mathcal{P}_{\textrm{min}}=N^{-K}$
since $\dim\mathcal{H}=N^{K}$. The second observation shall
be that if $\textrm{Tr}\left(\rho^{2}\right)\leq\mathcal{P}_{\textrm{PPT}}=\left(N^{K}-1\right)^{-1}$
then $\rho$ is PPT with respect to all possible partial transpositions
\cite{ball Multipartite,ball Bipartite}. This result is based on
the Mehta's theorem \cite{Mehta}.

To describe the ability to detect the NPPT states by the 
mixed--state collectibility we 
introduce parameters $\mathcal{P}_{N}^{\textrm{crit}}$
and $\mathcal{P}_{K}^{\textrm{crit}}$ related to inequalities (\ref{PPT inequality})
and (\ref{PPT inequality2}) respectively, and defined as follows.
If $\textrm{Tr}\left(\rho^{2}\right)\leq\mathcal{P}_{N}^{\textrm{crit}}$
or $\textrm{Tr}\left(\rho^{2}\right)\leq\mathcal{P}_{K}^{\textrm{crit}}$,
then the corresponding inequalities (\ref{PPT inequality}) or (\ref{PPT inequality2})
classify the state as PPT (even if it is NPPT). The critical values
of purity are the solutions to the equations:\begin{subequations}
\begin{equation}
r_{N}\left(N^{2},1,\mathcal{P}_{N}^{\textrm{crit}}\right)=N^{-2N},
\end{equation}
\begin{equation}
r_{2}\left(2^{K},1,\mathcal{P}_{K}^{\textrm{crit}}\right)=\frac{1}{16(2^{K-1}-1)}.
\end{equation}
\end{subequations}
This means, that we are looking for the values
of purity for which the general upper bound (\ref{nieoznaczonosc_mixed})
--- which increases with purity and tends to $0$ for the maximally
mixed state --- goes below two fixed values (depending on $N$ or
$K$) which appear on the right hand sides of (\ref{PPT inequality})
and (\ref{PPT inequality2}).

In Table \ref{Table1} we compared the critical values $\mathcal{P}_{N}^{\textrm{crit}}$
and $\mathcal{P}_{K}^{\textrm{crit}}$ with the general limitations
given by $\mathcal{P}_{\textrm{min}}$ and $\mathcal{P}_{\textrm{PPT}}$.
As it was expected both $\mathcal{P}_{N}^{\textrm{crit}}$ and $\mathcal{P}_{K}^{\textrm{crit}}$
are greater than $\mathcal{P}_{\textrm{PPT}}$, thus not all NPPT
states are detected by the criteria based on the inequalities (\ref{PPT inequality})
and (\ref{PPT inequality2}). However, in all cases the width of the
range of purities for which the NPPT states are not detected is less
than $0.05$, and seems to be slightly smaller for many qubits than
for two quNits. In both tables the second columns are the same because
they refer to the case of two qubits.

\begin{table} 
Bi--partite systems: $K=2$\qquad{}\qquad{}\qquad $K$ qubits: $N=2$ \qquad

\medskip{}

\begin{tabular}{|c|c|c|c|}
\hline 
$N$ & $2$ & $3$ & $4$\tabularnewline
\hline 
\hline 
$\mathcal{P}_{\textrm{min}}$ & 0.2500 & 0.1111 & 0.0625\tabularnewline
\hline 
$\mathcal{P}_{\textrm{PPT}}$ & 0.3333 & 0.1250 & 0.0667\tabularnewline
\hline 
$\mathcal{P}_{N}^{\textrm{crit}}$ & 0.3456 & 0.1728 & 0.1033\tabularnewline
\hline 
\end{tabular}\ \ %
\begin{tabular}{|c|c|c|c|}
\hline 
$K$ & $2$ & $3$ & $4$\tabularnewline
\hline 
\hline 
$\mathcal{P}_{\textrm{min}}$ & 0.2500 & 0.1250 & 0.0625\tabularnewline
\hline 
$\mathcal{P}_{\textrm{PPT}}$ & 0.3333 & 0.1429 & 0.0667\tabularnewline
\hline 
$\mathcal{P}_{K}^{\textrm{crit}}$ & 0.3456 & 0.1599 & 0.0808\tabularnewline
\hline 
\end{tabular}

\caption{Minimal purities for which the NPPT property is detected. The parameter
$\mathcal{P}_{\textrm{min}}$ denotes the minimal possible purity
of the system. The parameter $\mathcal{P}_{\textrm{PPT}}$ gives the
bound for the purity below which all states are PPT. The critical
parameters $\mathcal{P}_{N}^{\textrm{crit}}$ (left table; given as
a function of $N$, for $K=2$) and $\mathcal{P}_{K}^{\textrm{crit}}$
(right table; a function of $K$, for $N=2$) provide the bounds for
purities above which the NPPT property is detected by the criteria
(\ref{PPT inequality}) and (\ref{PPT inequality2}) respectively.}

\label{Table1}
\end{table}

\section{The bi--partite case}\label{sec:bipartiteWerner}

\subsection{The generalized Werner state}

We define a generalized Werner state on 
an $N \times N$ system as 
\begin{equation}
\rho_{w}=\alpha\left(U\otimes V\right)\ketbra{\psi_{\boldsymbol{\lambda}}}{\psi_{\boldsymbol{\lambda}}}\left(U\otimes V\right)^{\dagger}+\frac{1-\alpha}{N^{2}}\1,
\label{generalized_Werner}
\end{equation}
Here  $\left|\psi_{\boldsymbol{\lambda}}\right\rangle $ represents a normalized
pure state with the following Schmidt decomposition 
\begin{equation}
\left|\psi_{\boldsymbol{\lambda}}\right\rangle =\sum_{i=1}^{N}\sqrt{\lambda_{i}}\left|ii\right\rangle ,\qquad\sum_{i=1}^{N}\lambda_{i}=1,\label{wektor z lambdami}
\end{equation}
while $U\otimes V$ denotes a local unitary matrix. 
In Appendix \ref{App:Werner}
one can find a detailed derivation of the following 
expression for the mixed--state
collectibility of (\ref{generalized_Werner}): 
\begin{subequations}
\begin{equation}
Y^{{\rm max}}\left[\rho\right]=\alpha^{N-1}y\left(\boldsymbol{\lambda}\right)\left(\alpha+\frac{1-\alpha}{N^{2}}\left[y\left(\boldsymbol{\lambda}\right)\right]^{-1/N}\right),\label{Collectibility Werner}
\end{equation}
\begin{equation}
y\left(\boldsymbol{\lambda}\right)\equiv Y^{{\rm max}}\left[\left|\psi_{\boldsymbol{\lambda}}\right\rangle \right]=\left(\frac{\sum_{i}\sqrt{\lambda_{i}}}{N}\right)^{2N}.\label{Collectibility pure}
\end{equation}
\end{subequations} 
Since for $\alpha=1$ we have $Y^{{\rm max}}\left[\rho_{w}\right]=y\left(\boldsymbol{\lambda}\right)$,
the function $y\left(\boldsymbol{\lambda}\right)$ coincides 
with the pure--state collectibility of
$\left|\psi_{\boldsymbol{\lambda}}\right\rangle $ given by (\ref{wektor z lambdami}),
as emphasized in (\ref{Collectibility pure}). 
Note that $y\left(\boldsymbol{\lambda}\right)$ is a function of the
R{\'e}nyi entropy $H_{1/2}\left(\boldsymbol{\lambda}\right)$ of the Schmidt
vector $\boldsymbol{\lambda}=\left\{ \lambda_{1},\ldots,\lambda_{N}\right\} $,
as $Y^{{\rm max}}\left[\left|\psi_{\boldsymbol{\lambda}}\right\rangle \right]=N^{-2N}\exp\left[NH_{1/2}\left(\boldsymbol{\lambda}\right)\right]$.

In order to discuss the quality of the entanglement (PPT) criteria
(\ref{PPT inequality}) we shall examine an example of the generalized
Werner state (\ref{generalized_Werner}) of two qubits. In that case
we have only one Schmidt number $\lambda$, i.e.
\begin{equation}
\left|\psi_{\lambda}\right\rangle =\sqrt{\lambda}\left|00\right\rangle +\sqrt{1-\lambda}\left|11\right\rangle .
\end{equation}
If we apply the usual PPT criteria (in that case PPT property is equivalent
to separability) we find that the state (\ref{generalized_Werner})
is separable if $\alpha\leq\alpha_{T}=1/\left(1+4\omega\right)$,
where $\omega=\sqrt{\lambda\left(1-\lambda\right)}$. When the state
$\left|\psi_{\lambda}\right\rangle $ is separable, then $\rho_{w}$
is also always separable (for $\alpha\leq1$). In the opposite case,
when $\left|\psi_{\lambda}\right\rangle $ is maximally entangled,
then $\lambda=1-\lambda=1/2$ and the state $\rho_{w}$ is separable
for $\alpha\leq1/3$, as it shall be in the case of the original Werner
state \cite{werner}.

\begin{figure}
\includegraphics[scale=0.7]{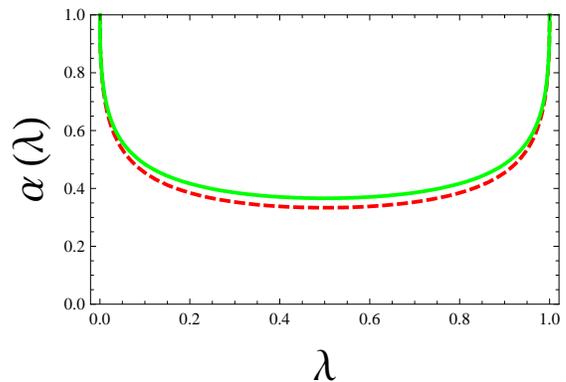}
\caption{(Color online). Separability of the generalized Werner state of two
qubits. Below the red, dashed line (representing $\alpha_{T}$) this
state is separable. Above the green line (representing $\alpha_{C}$)
our criteria (\ref{PPT inequality}) based on the mixed--state collectibility
detect its entanglement.}

\label{Fig:Werner Colectibility}
\end{figure}

We would like to specify, in which cases the mixed--state collectibility
given by the formula (\ref{Collectibility Werner}), is able to detect
entanglement of two--qubit generalized Werner state. This happens
when $Y^{{\rm max}}\left[\rho_{w}\right]>1/16$, thus, for 
\begin{equation}
\alpha>\alpha_{C}=\frac{2}{1+2\omega+\sqrt{\left(1+2\omega\right)\left(1+10\omega\right)}}.
\end{equation}
In Fig. \ref{Fig:Werner Colectibility} we compare two parameters
$\alpha_{T}$ and $\alpha_{C}$ as functions of $\lambda$. Both curves
lay close to each other, what shows that the mixed--state collectibility
provides very good efficiency of entanglement detection in the case
of the generalized Werner state of two qubits.

\subsection{Pure--state collectibility as a bipartite entanglement measure}

In the previous section we  calculated the mixed--state collectibility
(\ref{Collectibility Werner}) for the generalized Werner state (\ref{generalized_Werner}).
We showed that the formula (\ref{Collectibility Werner}) for the
value $\alpha=1$ is the pure--state collectibility of the bi--partite
state $\left|\psi_{\boldsymbol{\lambda}}\right\rangle $, i.e. $y\left(\boldsymbol{\lambda}\right)\equiv Y^{{\rm max}}\left[\left|\psi_{\boldsymbol{\lambda}}\right\rangle \right]$.
From that result we establish the following observation:
\begin{fact}
The pure--state collectibility of a bi--partite state $\left|\Psi\right\rangle \in\mathcal{H}_{2}$
is a function of the negativity 
\begin{equation}
Y^{{\rm max}}\left[\left|\Psi\right\rangle \right]=\frac{\left(1+\left(N-1\right)\mathcal{N}\left[\left|\Psi\right\rangle \right]\right)^{N}}{N^{2N}}.\label{Collectibility Werner-1}
\end{equation}
\end{fact} 
Negativity $\mathcal{N}$ is simple entanglement 
measure \cite{ball Bipartite,Negativity} which for a pure state reads
\begin{equation}
\mathcal{N}\left[\left|\Psi\right\rangle \right]=\frac{\left\Vert \left(\left|\Psi\right\rangle \left\langle \Psi\right|\right)^{T_{B}}\right\Vert _{1}-1}{N-1}.
\end{equation}

The above fact implies that the pure--state collectibility in the
bi--partite case can be alternatively defined in terms of the maximal
fidelity \cite{MaxFidelity1,MaxFidelity2}
\begin{equation}
Y^{{\rm max}}\left[\left|\Psi\right\rangle \right]=N^{-N}\max_{U,V}\left|\!\left.\left\langle \Psi\right|\Phi^{+}\right\rangle \right|^{2N},
\end{equation}
with respect to the maximally entangled state
\begin{equation}
\left|\Phi^{+}\right\rangle =\left(U\otimes V\right)\frac{1}{\sqrt{N}}\sum_{i=1}^{N}\left|ii\right\rangle .
\end{equation}
Maximization over the local unitaries $U\otimes V$ assures
that $Y^{{\rm max}}$ depends on the overlap of
the analyzed state $|\Psi\rangle$
 with the closest maximally entangled state. Thus, the pure--state collectibility can be considered as a quantity complementary to the geometric measure of entanglement \cite{geom1,geom2}. The latter describes the minimal distance of the state analyzed to the closest separable state, while the former is related to its distance to the closest maximally entangled state, i.e. the Bell state.

\section{Mixed--states entanglement criteria based on 
pure--state collectibility and remote purities}\label{sec:experiment2}

The mixed--states entanglement criteria (\ref{criteria_Col2}) are
provided by the inequality (\ref{corrected1}). In order to prove
this inequality we shall derive, using the purity parameters and the
separability assumption, the lower bound on the purity of the reduced
(unconditional) Bob state $\rho_{B}$. 

\subsection{Bound on total Bob purity via his conditional purities}

Consider a mixed state of two qubits $\rho_{AB}$ in a usual block
shape: 
\begin{equation}
\rho_{AB}=\left[\begin{array}{cl}
A_{+} & C\\
C^{\dagger} & A_{-}
\end{array}\right].\label{state}
\end{equation}
We may locally rotate the state to the form in which the block $A_{+}$ is diagonal.
Such an operation is performed on the Bob side, thus it commutes with the
von Neumann measurements on the side of Alice. 
We also put the following notation
\begin{equation}
A_{\pm}=p_{\pm}\sigma_{\pm},\label{matrices-states}
\end{equation}
where, as mentioned before, $p_{\pm}$ are probabilities of the results
{}``$\pm$'' of the von Neumann measurement of $\sigma_{\hat{z}}$
on system {}``$A$'' and $\sigma_{\pm}$ are the states created
remotely by Alice on Bob side after getting the results $\pm$.

Note that the axis $\hat{n}$ may be in general arbitrarily chosen
(here $\hat{n}=\hat{z}$). Let us write (\ref{state}) in a form of
the following $4\times4$ matrix: 
\begin{equation}
\rho_{AB}=\left[\begin{array}{clrr}
p_{+}p_{1} & w & c_{11} & c_{12}\\
w^{*} & p_{+}p_{2} & c_{21} & c_{22}\\
c_{11}^{*} & c_{21}^{*} & p_{-}q_{1} & z\\
c_{12}^{*} & c_{22}^{*} & z^{*} & p_{-}q_{2}
\end{array}\right].\label{state1}
\end{equation}
An analysis of the above structure leads immediately to the following\textit{
}lemma: 
\begin{lemma}
\label{LemH1}
Suppose that the state (\ref{state1}) is separable 
and in addition it satisfies the remote--purity condition (\ref{remote-purity}) 
under the measurement performed on system {}``$A$'' in basis $\hat{n}$.
Then we have: 

%If without loss of generality we shall assume 
%\begin{equation}
%p_1\geq p_2 
%\label{assumptions}
%\end{equation}

\begin{enumerate}
\item $p_{+}p_{-}p_{i}q_{j}\geq\left|c_{ij}\right|^{2}$ \quad for \; $i,j=1,2$ 
\item $p_{+}p_{-}p_{2}q_{1}\geq\left|c_{12}\right|^{2}$, $\qquad p_{+}p_{-}p_{1}q_{2}\geq\left|c_{21}\right|^{2}$
\item $p_{1}p_{2}\leq\epsilon_{+}/2$, $\qquad q_{1}q_{2}\leq\epsilon_{-}/2$ 
\end{enumerate}
\end{lemma}\proof The first statement follows from positivity of
$\rho_{AB}$, the second one comes from PPT condition equivalent to
separability, while the last one is implied by (\ref{remote-purity}).

% \footnote{Since (\ref{remote-purity}) is equivalent to $p_1^{2} + p_{2}^{2} \geq 1- \epsilon_{+}$,
%we get (iv) by exploiting $1=(p_1+p_2)^{2}=p_1^{2} +p_2^{2} + 2 p_1p_2$.
%The second part of (iv) comes form its first part and from the assumption (\ref{assumptions}): 
%$\frac{\epsilon}{2} \geq (p_1p_2)\geq (p_2)^{2}$.}}

Now we shall provide a key bound on the purity of the reduced state of Bob.

%{\it Fact 2a .-  The elements of submatrix $C$ of the density matrix form the Fact 1 satisfy the following restrictions:  
%\begin{eqnarray}
%&& |c_{11}|^{2} \leq p_+p_{-}p_1q_1 \leq p_{+}p_{-}  \\
%&& |c_{12}|^{2}, |c_{21}|^{2}, |c_{22}|^{2} \leq  p_+p_{-}p_2 \leq  p_{+}p_{-}\sqrt{\frac{\epsilon_+}{2}}.
%\label{C-ograniczenia}
%\end{eqnarray}}

\begin{lemma}
\label{LemH2} 
Suppose that the matrix from Lemma \ref{LemH1}
fulfills the remote--purity condition (\ref{remote-purity1}) under
the measurement along the perpendicular basis $\hat{n}'$,
 ($\hat{n}\cdot\hat{n}'=0$).
Then its reduced Bob density matrix, $\rho_{B}\equiv A_{+}+A_{-}$,
satisfies the purity condition: 
\begin{equation}
\textrm{Tr}\left(\rho_{B}^{2}\right)\geq1-\eta\left(\epsilon_{+}\epsilon_{-};p_{+}p_{-};\epsilon'\right),\label{bound}
\end{equation}
with $\epsilon'$ and $\eta$ introduced in Theorem \ref{th:bound1}.\end{lemma} 

A proof of Lemma \ref{LemH2} can be found in Appendix \ref{App:Experiment}.
Clearly the dual inequality obtained by swapping the roles of $\hat{n}$
and $\hat{n}'$ is also satisfied: 
\begin{equation}
\textrm{Tr}\left(\left(\rho_{B}'\right)^{2}\right)\geq1-\eta\left(\epsilon_{+}'\epsilon_{-}';p_{+}'p_{-}';\epsilon\right).\label{bound-dual}
\end{equation}
An intriguing property of formula (\ref{bound}) is that while
in the first basis $\hat{n}$ it requires only one conditional purity
to be small (since the product $\epsilon_{+}\epsilon_{-}$ is involved),
in the second basis $\hat{n}'$ it requires smallness of both purities, 
since \textit{$\epsilon'=\max\left[\epsilon_{+}',\epsilon_{-}'\right]$}.

\subsection{Experimentally suitable uncertainty relation for mixed two--qubit states}

Consider again the state $\rho_{B}$, obtained
as a reduction of the original bi--partite state (\ref{state1}).
We have the following immediate fact: \begin{fact}\label{LemH3} If
the purity of the state $\rho_{B}\equiv A_{+}+A_{-}$ with $A_{\pm}=p_{\pm}\sigma_{\pm}$
defined for some probabilities $\{p_{\pm}\}$ and states $\{\sigma_{\pm}\}$
satisfies $\textrm{Tr}\left(\rho_{B}^{2}\right)\geq1-\eta,$ then
\begin{equation}
\textrm{Tr}\left(A_{+}^{2}\right)+\textrm{Tr}\left(A_{-}^{2}\right)+2\textrm{Tr}\left(A_{+}A_{-}\right)\geq1-\eta.\label{bound-eta}
\end{equation}
\end{fact} Recall that the quantities 
in our interferometric scheme (see Fig. \ref{Fig.2}) 
 are \textit{directly measurable} under measuring both Alice qubits along
the direction $\hat{z}$. 
Equations (\ref{DefG1}) and (\ref{DefG2}) together with the
definition of $A_{\pm}$ imply the following extensions for the
expressions defined for the pure--state case:
\begin{equation}
G_{\pm\pm}:=\sqrt{\textrm{Tr}\left(A_{\pm}^{2}\right)},\qquad\left|G_{\pm\mp}\right|^{2}=\textrm{Tr}\left(A_{\pm}A_{\mp}\right).\label{definitions}
\end{equation}
In a full analogy we may define the quantities $G_{\pm\pm}'$ and
$G_{\pm\mp}'$ as putting the primes on RHS of the above equations
which means that all quantities were measured in the complementary
basis (say $\hat{x}$). Combining Lemma \ref{LemH2} and Fact \ref{LemH3}
we are immediately able to derive the central result:

\noindent{\it Proof of Theorem \ref{th:bound1}}. 
Directly from Eq. (\ref{definitions}) we have
\begin{equation}
G_{++}^{2}+G_{--}^{2}+2\left|G_{+-}\right|^{2}=\textrm{Tr}\left(A_{+}^{2}\right)+\textrm{Tr}\left(A_{-}^{2}\right)+2\textrm{Tr}\left(A_{+}A_{-}\right).\label{corrected1-1}
\end{equation}
Since we have assumed that the state $\rho_{AB}$ is separable, 
according to Lemma \ref{LemH2} we obtain  that $\textrm{Tr}\left(\rho_{B}^{2}\right)\geq1-\eta$.
Thus, the bound (\ref{bound-eta}) applied to Eq. (\ref{corrected1-1})
boils down after short rearrangement to the desired inequality (\ref{corrected1}).

Inequality (\ref{corrected1}) implies a bound for the mixed--states
generalization of the collectibility (\ref{Collectibility}):
\begin{equation}
Y_{a}\left[\rho_{AB}\right]\leq\frac{\left(\sqrt{2G_{++}G_{--}}+\sqrt{\eta+\left(G_{++}+G_{--}\right)^{2}-1}\right)^{2}}{8}.
\label{bound-fundamental}
\end{equation}
In order to derive the modified entanglement criteria (\ref{criteria_Col2})
we shall finally bound from above the terms $\sqrt{G_{++}G_{--}}$
and $G_{++}+G_{--}$ by $1/2$ and $1$ respectively.

The separable two--qubit state $\rho_{AB}$ also satisfies the dual
inequalities generated automatically by interchanging the roles of
the two complementary bases. Such operation is equivalent to putting
in Eq. (\ref{bound-fundamental}) prime at all the quantities (particularly
in the arguments of the $\eta$ function) other than $\epsilon'$
and turning the latter into $\epsilon=\textrm{max}\left[\epsilon_{+},\epsilon_{-}\right]$.

\subsection{Discussion and simplifications}

Let us discuss the inequality (\ref{bound-fundamental}) for 
three elementary examples.
 First of all, if the state is pure and separable we have $\eta=0$
and $G_{++}+G_{--}=1$ as any separable pure state is a product state. 
This implies that all elements of the Gram matrix  $G$ are equal
so the inequality (\ref{bound-fundamental}) is saturated.
Secondly, if a separable mixed state has a product structure
then the coefficients {}``$G$''
are all equal, so that the left hand side of this inequality reads
 $Y_{a}\left[\rho_{AB}\right]=G_{++}G_{--}/4$.
Since the state is mixed there is a contribution of a positive parameter $\eta>0$ 
on the right hand side so in this case the inequality holds and it is strict.
 Finally, if  the state is pure but maximally entangled then $\eta=0$ so 
the right hand side of (\ref{bound-fundamental}) is equal to $1/16$ 
while the collectibility on the left side is equal to $1/4$,
as shown earlier in  \cite{Collectibility_RHZ},
so the inequality is violated by a factor of four.

Let us emphasize that the performed analysis leading to (\ref{bound-fundamental})
is independent of the choice of the two complementary observables,
so we may consider an \textit{optimization of the two dual inequalities}
over two measurements of the two complementary observables $\{\hat{n}\vec{\sigma},\hat{n}'\vec{\sigma}\}$
with the constraint $\hat{n}\cdot\hat{n}'=0$.

Furthermore, let us observe that  inequality (\ref{bound-fundamental})
is a slightly stronger variant of the original result $Y_{a}\left[\left|\Psi\right\rangle _{AB}\right]\leq1/16$
valid for separable pure states. Since the state is pure we shall
put $\eta=0$ and $G_{++}+G_{--}=1$ in (\ref{bound-fundamental}),
and obtain:
\begin{equation}
Y_{a}\left[\left|\Psi\right\rangle _{AB}\right]\leq\frac{G_{++}G_{--}}{4}\leq\frac{1}{16}.\label{original-stronger}
\end{equation}
The above inequality is useful for entanglement detection in the case
of pure states that are not maximally entangled.

Let us make a remark on  statistical properties of these inequalities.
Inequality (\ref{bound-fundamental}), as well as the separability
criteria based on $Y_{a}\left[\rho_{AB}\right]$, involve the square
root calculated on the difference of the measurable quantities in
the form $\sqrt{G_{++}G_{--}-\left|G_{+-}\right|^{2}}$. While in
theory the argument inside the root is always positive, in practice,
due to experimental errors, a problems with the sign under the square root
might appear. We shall keep in mind that the violation of the inequality is equivalent
to ,,passing'' the entanglement test. The most natural approach is
to use the term ,,passed'' to all cases where the quantity under
the square root in LHS of (\ref{bound-fundamental}) is strictly positive
up to the error bars. This is actually the test of the quantity of the
entanglement produced since in all entangled states the term $G_{++}G_{--}-\left|G_{+-}\right|^{2}$
becomes positive, sometimes in a very drastic way. For instance,
 for maximally entangled states the second term $\left|G_{+-}\right|$
 vanishes, and $G_{++}G_{--}=1/4$.
Therefore,  testing the degree of entanglement of the state analyzed,
we look for the case in which the number $G_{++}G_{--}-\left|G_{+-}\right|^{2}$ is significantly positive. This justifies  the term ,,violated''
or the term ,,passed'' from the perspective of testing the entanglement property.

Finally, let us point out that other separability tests, based on 
a pair of inequalities (\ref{bound}) and (\ref{bound-dual}) 
or  (\ref{corrected1}) and its dual, do not involve the problem
discussed above. In fact, these inequalities seem to be more suitable
for an experimental application.
In this context the previous paragraph dealing with the issue 
of the square may seem to be of purely academic character. We keep it since 
the reasoning as it is has an element of the universality --- in 
all the cases when the error may spoil the mathematics 
of the test formula one should look at the region of 
parameters in which the test really matters.

\section{Conclusions}

In this work we generalized the collectibility --- a quantity initially designed to
characterize entanglement of pure quantum states --- for mixed states. 
On one hand we improved existing expressions for pure--state collectibility
by taking into account contributions due to non--maximal purity 
of the state investigated. This approach can be useful in practice
to analyze experimentally states intended to be pure, which are characterized by a high purity. For the simplest case of two qubit system
the proposed measurement scheme requires two copies of the  state analyzed,
so a possible optical setup involves four photon experiments.

For such pseudo--mixed states 
the essential element of the scheme is that one performs two measurements represented
by the orthonormal Bloch vectors $\hat{n}$ and $\hat{n}'$
corresponding to the two polarization rotations
$R^{\dagger}(\theta,\phi)$, and $R^{\dagger}(\theta',\phi')$.
This allows to avoid the assumption which was necessary in \cite{Collectibility_RHZ}, that the state is pure.

On the other hand we introduced here a new notion of the mixed--state collectibility ---
a quantity defined for an arbitrary mixed state of a composite system 
which contains $K$ subsystems, each describing an $N$--level system. 
Presence of the $N$--th root in the definition (\ref{MCm})
assures that it is consistent with the former notion of 
pure--state collectibility (\ref{MCp}). In the particular
case of a pure state of a bipartite system 
the collectibility is shown to be a function of the negativity.
Moreover, in this case the collectibility depends on the minimal 
distance of the pure state analyzed to the closest maximally entangled state.

Explicit bounds for the mixed--state collectibility obtained for arbitrary
quantum states, separable states and states with positive partial transpose
belong to the key results of the present paper. They allow us to
design practical tests for entanglement of a given state.
Such experimental schemes look realistic at least for the mixed state of a 
two qubit system, for which a four--photon experiment is necessary.
We are thus tempted to believe that such experiments, useful to demonstrate 
quantum entanglement of a given state without performing its quantum tomography,
will be realized in a near future.

There are still few  more questions for future research.
One should explore possible structural connections of the present
quantity to the symmetric measurement state reconstruction of
the type proposed in Ref. \cite{MoroderEtAl2012} and to the
experimental discrimination of SLOCC--invariant classes
\cite{OsterlohSiewert2012}.
The directly related problem is the issue of
lower bounds for typical entanglement measures (like concurrence or the
geometric measure) based on the measurement reproducing
collectibility in analogy to the standard approach
(see \cite{RMPH1999,EisertEtAl2007,GuehneEtAl2006,AudenaertPlenio2006}).
One of the interesting practical questions 
that will be considered elsewhere is what are
the best possible lower bounds on entanglement measures
at the constraints given by full output data from
the presented Hong--Ou--Mandel setup.  While the issue of finding
some  lower bounds for the latter (two--qubit) case is relatively easily
tractable the analogous questions for higher dimensions
and/or subsystems seem to require much more effort.
Finally, there is an intriguing question of the relation 
of the test designed for pseudopure entangled states 
to the entanglement criteria based on mutually unbiased bases (MUB--s)
provided in Ref. \cite{SpenglerEtAl2012}. In fact, the H--O--M based test for pseudopure qubit entanglement uses incomplete 
MUB in one lab being two Pauli matrices. However one should 
be careful with the way the MUB--s are used here
in order to keep  the parametric efficiency of the scheme in context 
of direct tomography method.

\begin{acknowledgments}
It is a pleasure to thank Florian Mintert,
Otfried G{\"u}hne, G{\'e}za T{\'o}th and Jens Siewert 
%for discussions.
%%%% who else???
%%%%%%%%%%%%%%%%%%%%%
 for fruitful discussions and helpful remarks.
This research was supported by the grants number: IP2011 046871
({\L}.R.),  N N202-261--938 (P.H.)
and   N N202--090-239  (K.{\.Z}.)
%%%% Zbyszek - your PIN number??
%%%%%%%%%%%%%%%%%%%%%
of the Polish Ministry of Science and Higher Education. A partial support from EC through the project Q--ESSENCE (P.H.) and by a postdoc internship (Z.P.), decision number DEC--2012/04/S/ST6/00400, from the Polish National Science Centre are gratefully acknowledged.

\end{acknowledgments}
\appendix\section{The maximum of a product of matrix elements}\label{App:product}

\noindent{\it Proof of Lemma \ref{lemma1}}. 
We shall maximize $\prod_{i,j=1}^{N}\rho_{ij}$ over all possible
choices of the matrix entries $\rho_{ij}$ of a hermitian $D\times D$
matrix $\rho$, assuming two constraints: $\textrm{Tr}\rho=\textrm{const}$
and $\textrm{Tr}\left(\rho^{2}\right)=\textrm{const}$. This means
to solve the following system of equations (for all $k,m=1,\ldots,D$):
\begin{equation}
\frac{\partial}{\partial\rho_{km}}\left[\prod_{i,j=1}^{N}\rho_{ij}-\sum_{i,j=1}^{D}\rho_{ij}\left(\nu\delta_{ij}+\frac{\mu}{2}\rho_{ji}\right)\right]=0,
\end{equation}
where $\nu$ and $\mu/2$ are the Lagrange multipliers related to
the constraints on the trace and purity respectively. After differentiation
with respect to $\rho_{km}$ we obtain\begin{subequations}
\begin{eqnarray}
C\rho_{km}^{-1} & = & \nu\delta_{km}+\mu\rho_{mk}\qquad k,m\leq N,\label{AppAeq1}\\
0 & = & \nu\delta_{km}+\mu\rho_{mk}\qquad\textrm{elsewhere},\label{AppAeq2}
\end{eqnarray}
\end{subequations}where $0\leq C\equiv\prod_{i,j=1}^{N}\rho_{ij}$
shall be from now on treated as a constant. Note that $\rho_{km}^{-1}\equiv1/\rho_{km}$.
In order to prove Lemma \ref{lemma1} we have to find the maximal value of $C$. 

The structure of the above equations provides that the maximizing
matrix $\rho$ will have the form: 
\begin{equation}
\rho_{km}=\textrm{Tr}\rho\begin{cases}
\rho_{a} & k,m\leq N\;\wedge\; k=m\\
\rho_{b}\, e^{i\varphi_{km}} & k,m\leq N\;\wedge\; k\neq m\\
\rho_{c} & k,m>N\;\wedge\; k=m\\
0 & k,m>N\;\wedge\; k\neq m
\end{cases},\label{lem1_solution}
\end{equation}
determined by three real coefficients $\rho_{a}$, $\rho_{b}\geq0$,
$\rho_{c}$ and all phases $\varphi_{km}$ being arbitrary. For further
convenience we have separated the normalization
factor $\textrm{Tr}\rho$.

The equations (\ref{AppAeq1}, \ref{AppAeq2}) together with both
constraints give five conditions:\begin{subequations}
\begin{equation}
C\left(\rho_{a}\textrm{Tr}\rho\right)^{-1}=\nu+\mu\rho_{a}\textrm{Tr}\rho,\label{system1}
\end{equation}
\begin{equation}
C\left(\rho_{b}\textrm{Tr}\rho\right)^{-1}=\mu\rho_{b}\textrm{Tr}\rho,\label{system2}
\end{equation}
\begin{equation}
\nu+\mu\rho_{c}\textrm{Tr}\rho=0,\label{system3}
\end{equation}
\begin{equation}
N\rho_{a}+\left(D-N\right)\rho_{c}=1,\label{system4}
\end{equation}
\begin{equation}
N\rho_{a}^{2}+N\left(N-1\right)\rho_{b}^{2}+\left(D-N\right)\rho_{c}^{2}=\xi,
\end{equation}
where $\xi=\textrm{Tr}\left(\rho^{2}\right)/\left(\textrm{Tr}\rho\right)^{2}$,
so that $\xi\in\left[D^{-1},1\right]$. In order to solve the above
system of equations we shall rewrite Eqs. (\ref{system2}, \ref{system3})
to the form $\left(\rho_{b}\textrm{Tr}\rho\right)^{2}=C/\mu$ and
$\rho_{c}\textrm{Tr}\rho=-\nu/\mu$, and eliminate both Lagrange multipliers
from (\ref{system1}) to obtain
\begin{equation}
\rho_{b}^{2}\rho_{a}^{-1}+\rho_{c}-\rho_{a}=0.\label{systemplus}
\end{equation}
\end{subequations} From Eqs. (\ref{system4}--\ref{systemplus})
we find (we introduce the dual dimension $\tilde{D}=D-N$):\begin{subequations}
\begin{equation}
\rho_{a}=\frac{1-\tilde{D}\rho_{c}}{N},\qquad \rho_{b}=\frac{\sqrt{\left(1-D\rho_{c}\right)\left(1-\tilde{D}\rho_{c}\right)}}{N},
\end{equation}
and the quadratic equation for $\rho_{c}$ 
\begin{equation}
D\tilde{D}\rho_{c}^{2}-\left(D+\tilde{D}-1\right)\rho_{c}+1-\xi=0.
\end{equation}
\end{subequations} The above equation possesses two non--negative
solutions of the form 
\begin{equation}
\rho_{c}^{\pm}=\frac{D+\tilde{D}-1\pm\sqrt{\left(D+\tilde{D}-1\right)^{2}+4D\tilde{D}\left(\xi-1\right)}}{2D\tilde{D}},
\end{equation}
that lead to two independent solutions of the main problem:
\begin{equation}
C_{\pm}=\left(\frac{\textrm{Tr}\rho}{N}\right)^{N^{2}}\left(1-D\rho_{c}^{\pm}\right)^{\frac{N\left(N-1\right)}{2}}\left(1-\tilde{D}\rho_{c}^{\pm}\right)^{\frac{N\left(N+1\right)}{2}}.
\end{equation}
Since $\rho_{c}^{-}\leq\rho_{c}^{+}$, the larger value is always given
by $C_{-}$ (this statement holds also when $1-D\rho_{c}^{+}$ and/or
$1-\tilde{D}\rho_{c}^{+}$ become negative). The above conclusion
finalizes the proof of Lemma \ref{lemma1} --- we shall only rename $\left[C_{-}\right]^{1/N}$
by $r_{N}\left(D,\textrm{Tr}\rho,\textrm{Tr}\left(\rho^{2}\right)\right)$. 
It is important to point out that the solution presented above provides a global maximum, since the set of $D\times D$ hermitian matrices with given trace and Hilbert--Schmidt norm (which coincides with purity in the case of positive semi--definite matrices) has topology of $D^2-2$ sphere $\mathcal{S}^{D^2-2}$, so that it has no boundary. 

\section{Collectibility of the generalized Werner state}
\label{App:Werner}

Before we start calculating the mixed--state collectibility for the
generalized Werner state (\ref{generalized_Werner}) we need the following
lemma
\begin{lemma}
\label{lemma2}
For $b\geq0$ and $q\geq1$ the
function $h:\left[0,\infty\right)^{N}\to\left[0,\infty\right)$ 
\begin{equation}
h\left(\boldsymbol{x}\right)=\prod_{i=1}^{N}\left(x_{i}^{2}+b\right)x_{i}^{q},
\end{equation}
is Schur concave.

\end{lemma}\proof Since $x_{i}\geq0$, the Theorem~II.3.14 from
\cite{bhatia1997matrix} states that to prove the Schur
concavity of the function $h\left(\boldsymbol{x}\right)$, it is sufficient
to show that $\Theta\left[h\left(\boldsymbol{x}\right)\right]\leq0$,
where:
\begin{equation}
\Theta\left[h\left(\boldsymbol{x}\right)\right]\equiv\left(x_{j}-x_{k}\right)\left(\frac{\partial h}{\partial x_{j}}\left(\boldsymbol{x}\right)-\frac{\partial h}{\partial x_{k}}\left(\boldsymbol{x}\right)\right).\label{Theta}
\end{equation}
An explicit computation gives
\begin{align}
\Theta\left[h\left(\boldsymbol{x}\right)\right]=-\left(x_{j}-x_{k}\right)^{2}\left(x_{j}x_{k}\right)^{q-1}\prod_{i\notin\{j,k\}}\left(x_{i}^{2}+b\right)x_{i}^{q}\nonumber \\
\times\left[b_{1}x_{j}^{2}x_{k}^{2}+b_{2}\left(x_{k}^{2}+x_{j}^{2}\right)+b\left(x_{k}-x_{j}\right)^{2}+qb^{2}\right] & ,
\end{align}
where $b_{1}=q+2$ and $b_{2}=b\left(q-1\right)$. Since $b_{1}\geq0$
and $b_{2}\geq0$ we obtain that $\Theta\left[h\left(\boldsymbol{x}\right)\right]\leq0$,
thus $h(\boldsymbol{x})$ is Schur concave.

The maximally mixed part of the Werner state (\ref{generalized_Werner})
is invariant under local unitary operations. Since the definition
(\ref{wektor z lambdami}) of $\left|\psi_{\boldsymbol{\lambda}}\right\rangle $
involves $U\otimes V$, we can choose the
basis of $N$ separable states to be $|\chi_{j}^{sep}\rangle=|jj\rangle$
without losing the generality.
The optimization over $|\chi^{sep}\rangle$ shall be then substituted
by maximization over $U\otimes V$, where
\begin{equation}
U=\sum_{m,n}u_{mn}\left|m\right\rangle \left\langle n\right|,\qquad V=\sum_{\mu,\nu}v_{\mu\nu}\left|\mu\right\rangle \left\langle \nu\right|.
\end{equation}

An explicit calculation yields
\begin{equation}
\left\langle \chi_{j}^{sep}\right|\rho_{w}\left|\chi_{k}^{sep}\right\rangle =\alpha g_{j}g_{k}^{*}+\frac{1-\alpha}{N^{2}}\delta_{jk},\label{explicit}
\end{equation}
where
\begin{equation}
g_{m}=\sum_{n=1}^{N}u_{mn}v_{mn}\sqrt{\lambda_{n}}.
\end{equation}
Now we shall use the result (\ref{explicit}) to calculate the product
appearing inside the definition (\ref{MCm})
\begin{equation}
\!\prod_{j,k=1}^{N}\!\!\!\left\langle \chi_{j}^{sep}\right|\!\rho\left|\chi_{k}^{sep}\right\rangle \!=\!\prod_{i=1}^{N}\!\!\left[\!\left(\!\sqrt{\alpha}\left|g_{i}\right|\right)^{2}\!+\frac{1\!-\!\alpha}{N^{2}}\!\right]\!\!\left(\sqrt{\alpha}\left|g_{i}\right|\right)^{2\left(N-1\right)}\!\!\!.\label{Product1}
\end{equation}
The right hand side of (\ref{Product1}) might be viewed as the $h(\boldsymbol{x})$
function from Lemma \ref{lemma2} with $x_{i}=\sqrt{\alpha}\left|g_{i}\right|$,
$b=\left(1-\alpha\right)/N^{2}$ and $q=2\left(N-1\right)$. 

Every vector $\boldsymbol{x}\in[0,\infty)^{N}$ majorizes the vector
of mean values $\bar{\boldsymbol{x}}=\left(\bar{x}_{1},\bar{x}_{2},\ldots,\bar{x}_{N}\right)$,
such that $\bar{x}_{1}=\bar{x}_{2}=\ldots=\bar{x}_{N}\equiv N^{-1}\sum_{j=1}^{N}x_{j}$.
According to Lemma \ref{lemma2}, $h(\boldsymbol{x})$ is the Schur
concave function, thus since $\boldsymbol{x}\succ\bar{\boldsymbol{x}}$
we obtain that $h(\boldsymbol{x})\leq h(\bar{\boldsymbol{x}})$. This
means that in each factor (for each $i$) inside the product (\ref{Product1})
we can substitute $\sqrt{\alpha}\left|g_{i}\right|$ by the number
$\sqrt{\alpha}N^{-1}\sum_{j=1}^{N}\left|g_{j}\right|$.
 Since we obtain then the $N$--th power of the same factor, 
 after some simplifications we find that
\begin{equation}
\prod_{j,k=1}^{N}\!\!\!\left\langle \chi_{j}^{sep}\right|\!\rho\left|\chi_{k}^{sep}\right\rangle \leq\left[\alpha^{N-1}p\left(\boldsymbol{\lambda}\right)\!\!\left(\!\alpha+\frac{1-\alpha}{N^{2}}\left[p\left(\boldsymbol{\lambda}\right)\right]^{-\frac{1}{N}}
\right)\!\!\right]^{N}\!\!\! .
\label{AppB Semifinal}
\end{equation}
We introduced  here the function  $p\left(\boldsymbol{\lambda}\right)$, 
\begin{equation}
p\left(\boldsymbol{\lambda}\right)=\left(\frac{1}{N}\sum_{i=1}^{N}\left|g_{i}\right|\right)^{2N}\!\!=\left(\frac{1}{N}\sum_{i=1}^{N}\left|\sum_{n=1}^{N}u_{in}v_{in}\sqrt{\lambda_{n}}\right|\right)^{2N}\!\!.
\end{equation}

Since the right hand side of (\ref{AppB Semifinal}) is an increasing
function of $p\left(\boldsymbol{\lambda}\right)$ we will prove the
result (\ref{Collectibility Werner}) showing that $p\left(\boldsymbol{\lambda}\right)\leq y\left(\boldsymbol{\lambda}\right)$.
To this end we establish the chain of inequalities:

\begin{eqnarray}
p\left(\boldsymbol{\lambda}\right) & \leq & \left[\frac{1}{N}\sum_{i=1}^{N}\sum_{n=1}^{N}\left|u_{in}\right|\left|v_{in}\right|\sqrt{\lambda_{n}}\right]^{2N}\\
 & = & \left[\frac{1}{N}\sum_{n=1}^{N}\sqrt{\lambda_{n}}\left(\sum_{i=1}^{N}\left|u_{in}\right|\left|v_{in}\right|\right)\right]^{2N}\nonumber \\
 & \leq & \left[\frac{1}{N}\sum_{n=1}^{N}\sqrt{\lambda_{n}}\sqrt{\sum_{i=1}^{N}\left|u_{in}\right|^{2}}\sqrt{\sum_{i=1}^{N}\left|v_{im}\right|^{2}}\right]^{2N}\nonumber \\
 & = & y\left(\boldsymbol{\lambda}\right).\nonumber 
\end{eqnarray}
The first one is the inequality for the modulus of a sum which in
the simplest form reads $\left|a+b\right|\leq\left|a\right|+\left|b\right|$.
Then we change the order of summation and apply the Cauchy--Schwarz
inequality for the sum over $i$. In the last step we use the fact
that the matrices $U$ and $V$ are unitary, 
so that $\sum_{i=1}^{N}\left|u_{in}\right|^{2}=1$
and $\sum_{i=1}^{N}\left|v_{im}\right|^{2}=1$
for any index $n$ and $m$.

Taking the $N$--th root of the expression (\ref{AppB Semifinal})
and applying the inequality $p\left(\boldsymbol{\lambda}\right)\leq y\left(\boldsymbol{\lambda}\right)$
we find that 
\begin{equation}
Y^{{\rm max}}\left[\rho_{w}\right]\leq\alpha^{N-1}y\left(\boldsymbol{\lambda}\right)\left(\alpha+\frac{1-\alpha}{N^{2}}\left[y\left(\boldsymbol{\lambda}\right)\right]^{-1/N}\right).
\label{prelast step}
\end{equation}
The last step is to show that the above inequality can be saturated.
Comparing (\ref{Product1}) with the inequality (\ref{prelast step})
we observe that this can happen only if there exist unitary matrices
$U$ and $V$ such that $\left|g_{i}\right|=\left[y\left(\boldsymbol{\lambda}\right)\right]^{1/\left(2N\right)}$
for all $i$. 
In that case the matrix elements $u_{in}$ and $v_{in}$
need to fulfill $N$ conditions of the form 
\begin{equation}
\forall_{i}\qquad\left|\sum_{n=1}^{N}u_{in}v_{in}\sqrt{\lambda_{n}}\right|=\frac{1}{N}\sum_{n=1}^{N}\sqrt{\lambda_{n}}.\label{satur}
\end{equation}
The solution of (\ref{satur}) might be constructed as $\left|u_{in}\right|=1/\sqrt{N}$
and $v_{in}=u_{in}^{*}$. Finally, since  inequality (\ref{prelast step})
can be saturated, the mixed--state collectibility for the generalized
Werner state (\ref{generalized_Werner}) is given by (\ref{Collectibility Werner}).

\section{Properties of Bob state purity in pseudopure two--qubit state}\label{App:Experiment}

\noindent{\it Proof of Lemma \ref{LemH2}}. After the measurement
performed on the subsystem {}``$A$'' in a basis complementary to
$\hat{z}$ (let's say $\hat{x}$) the matrices $A_{\pm}'$ (in the
second basis) are

\begin{equation}
A_{\pm}'\equiv p_{\pm}'\sigma_{\pm}'=\frac{1}{2}\left[\rho_{B}\pm X\right],\label{bloki-nowe}
\end{equation}
where $\rho_{B}=A_{+}+A_{-}$ and $X=C+C^{\dagger}$. Related normalization
constants read: 
\begin{equation}
p_{\pm}'\equiv\textrm{Tr}\left(A_{\pm}'\right)=\frac{1\pm\textrm{Tr}X}{2}.\label{p-primed}
\end{equation}
Note that the last term in (\ref{bloki-nowe}) is the most important
since we are looking for a restriction on the purity of $\rho_{B}$. 

We shall now consider the remote--purity condition (\ref{remote-purity1})
together with the Eqs. (\ref{bloki-nowe}) and (\ref{p-primed}) %Latter we shall take $\epsilon' = max \epsilon_{+}', \epsilon_{-}'$.
to obtain two inequalities (for both signs): 
\begin{equation}
\textrm{Tr}\left[\left(\rho_{B}\pm X\right)^{2}\right]\geq\left(1\pm\textrm{Tr}X\right)^{2}\left(1-\epsilon_{\pm}'\right),
\end{equation}
We extract the purity of the state $\rho_{B}$ as follows:
\begin{align}
\textrm{Tr}\left(\rho_{B}^{2}\right)\geq & \left(1-\epsilon_{\pm}'\right)\left(1+\left(\textrm{Tr}X\right)^{2}\right)-\textrm{Tr}\left(X^{2}\right)\nonumber \\
 & \pm2\left(1-\epsilon_{\pm}'\right)\textrm{Tr}X\mp2\textrm{Tr}\left(\rho_{B}X\right),
\end{align}
In order to eliminate the term $2\textrm{Tr}\left(\rho_{B}X\right)$
we shall add the inequalities of both signs and divide the result
by $2$
\begin{equation}
\textrm{Tr}\left(\rho_{B}^{2}\right)\geq\left(1-\bar{\epsilon}'\right)\left(1+\left(\textrm{Tr}X\right)^{2}\right)-\Delta'\textrm{Tr}X-\textrm{Tr}\left(X^{2}\right),
\end{equation}
 where $\bar{\epsilon}'=\left(\epsilon_{+}'+\epsilon_{-}'\right)/2$
and $\Delta'=\epsilon_{+}'-\epsilon_{-}'$. Next, we use the
identity $2\det X=\left(\textrm{Tr}X\right)^{2}-\textrm{Tr}\left(X^{2}\right)$
valid for $2\times2$ matrices to rewrite
\begin{equation}
\textrm{Tr}\left(\rho_{B}^{2}\right)\geq1-\bar{\epsilon}'+2\det X-\bar{\epsilon}'\left(\textrm{Tr}X\right)^{2}-\Delta'\textrm{Tr}X.
\end{equation}
Since the sign of $\det X$ is not fixed we shall involve an estimation
$\det X\geq-\left|\det X\right|$ which becomes an equality whenever
the sign of $\det X$ is negative. We obtain:
\begin{equation}
\textrm{Tr}\left(\rho_{B}^{2}\right)\geq1-\left(\bar{\epsilon}'+2\left|\det X\right|+\bar{\epsilon}'\left(\textrm{Tr}X\right)^{2}+\Delta'\textrm{Tr}X\right),
\end{equation}
thus, in the next step we shall maximize the following expression:
\begin{equation}
\bar{\epsilon}'+2\left|\det X\right|+\bar{\epsilon}'\left(\textrm{Tr}X\right)^{2}+\Delta'\textrm{Tr}X.
\end{equation}

Since 
\begin{equation}
X=\left[\begin{array}{cl}
2\textrm{Re}c_{11} & c_{12}+c_{21}^{*}\\
c_{12}^{*}+c_{21} & 2\textrm{Re}c_{22}
\end{array}\right],\label{state-1}
\end{equation}
we estimate the modulus of the determinant as:
\begin{eqnarray}
\left|\det X\right| & = & \left|4\textrm{Re}c_{11}\textrm{Re}c_{22}-\left|c_{12}+c_{21}^{*}\right|^{2}\right|\nonumber \\
 & \leq & 4\left|c_{11}\right|\left|c_{22}\right|+\left(\left|c_{12}\right|+\left|c_{21}\right|\right)^{2}.
\end{eqnarray}
Multiplying different conditions from Lemma \ref{LemH1} we find the
following estimates (only in this place we use the assumption that
$\rho_{AB}$ is separable): 
\begin{equation}
\left|c_{ij}\right|^{2}\leq p_{+}p_{-}\frac{\sqrt{\epsilon_{+}\epsilon_{-}}}{2},\qquad i\neq j,
\end{equation}
\begin{equation}
\left|c_{11}\right|\left|c_{22}\right|\leq p_{+}p_{-}\frac{\sqrt{\epsilon_{+}\epsilon_{-}}}{2},
\end{equation}
what imply 
\begin{equation}
\left|\det X\right|\leq4p_{+}p_{-}\sqrt{\epsilon_{+}\epsilon_{-}}.\label{det}
\end{equation}

In order to estimate the quadratic expression we observe, that according
to (\ref{p-primed}) we have $\textrm{Tr}X\in\left[-1,1\right]$.
Moreover, since $\bar{\epsilon}'\geq0$ and $\left|\Delta'\right|\leq1/2$,
there are two cases. For $\Delta'\geq0$ the maximum is attained
for $\textrm{Tr}X=1$, while for $\Delta'\leq0$ the maximal
value is provided by $\textrm{Tr}X=-1$. This implies the following
estimate
\begin{equation}
\bar{\epsilon}'\left(\textrm{Tr}X\right)^{2}+\Delta'\textrm{Tr}X\leq\bar{\epsilon}'+\left|\Delta'\right|.\label{binomial}
\end{equation}

Eventually, inequalities (\ref{det}) and (\ref{binomial}) give
the result
\begin{equation}
\bar{\epsilon}'+2\left|\det X\right|+\bar{\epsilon}'\left(\textrm{Tr}X\right)^{2}+\Delta'\textrm{Tr}X\leq8p_{+}p_{-}\sqrt{\epsilon_{+}\epsilon_{-}}+2\epsilon'\equiv\eta,
\end{equation}
where $\epsilon'=\max\left[\epsilon_{+}',\epsilon_{-}'\right]$. In
the last step we used the identity $\epsilon_{+}'+\epsilon_{-}'+\left|\epsilon_{+}'-\epsilon_{-}'\right|=2\epsilon'$.

\end{document}